\documentstyle[12pt]{article}
\def\be{\nopagebreak[3]\begin{equation}}
\newcommand{\ee}{\end{equation}}
\def\ba{\begin{array}}
\def\ea{\end{array}}

\def\ni{\noindent}
\newcommand{\scs}[1]{{\scriptscriptstyle #1}}
\newcommand{\F}{{\scriptscriptstyle F}}
\renewcommand{\Im}{{\rm Im}\,}
\renewcommand{\Re}{{\rm Re}\,}
\newcommand{\pvint}{\,-\!\!\!\!\!\!\int}
\begin{document}
\begin{titlepage}
\begin{flushright}
NBI-HE-92-78\\
A-204-1192\\
November 1992\\
Revised
\end{flushright}
\vspace*{36pt}
\begin{center}
{\huge \bf
Infinite-tension strings at $d>1$}
\end{center}
\vspace{2pc}
\begin{center}
 {\Large D.V. Boulatov}\\
\vspace{1pc}
{\em The Niels Bohr Institute\\
University of Copenhagen\\
Blegdamsvej 17, 2100 Copenhagen \O \\
Denmark\\
{\rm and}\\
Centre de Physique Theorique, Ecole Polytechnique\\
91128 Palaiseau, France}
\vspace{2pc}
\end{center}
\begin{center}
{\large\bf Abstract}
\end{center}
A matrix model describing surfaces embedded in a Bethe lattice is considered.
{}From the mean field point of view, it is equivalent to the
Kazakov-Migdal induced gauge theory and therefore, at $N=\infty$ and $d>1$,
the latter can be interpreted as a matrix model for infinite-tension strings.
We show that, in the naive continuum limit, it is governed by the
one-matrix-model saddle point with an upside-down potential.
To derive mean field equations, we consider the one-matrix model
in external field. As a simple application, its
explicit solution in the case of the inverted W potential is given.

\vfill
\end{titlepage}

\newpage

\section{Introduction}

Recently the ``induced QCD'' model  proposed by V.Kazakov and A.Migdal
\cite{KM} has drawn much attention \cite{Mig,Kogan,Mig2,Gross}.
This model possesses the local $U(1)^{\otimes N}$ symmetry
reminiscent to matrix models \cite{Boul} which makes it inequivalent to QCD.
Actually, to see that, it is sufficient to consider only
the discrete $Z_N$ subgroup which leads to the local-confinement
selection rule \cite{Kogan} incompatible with the usual QCD area law.

However, as will be shown
in the present paper, the Kazakov-Migdal (KM) model is interesting by itself.
In the planar limit it is soluble by the mean field technique. But, within
this approximation, it is equivalent to a matrix model
with a Bethe-lattice (BL) embedding
space. It means that all possible configurations are tree-like, {\em i.e.},
have no embedding area.
Therefore, it is natural to consider the KM model
as an infinite-tension limit of a $d$-dimensional matrix model.
Indeed, in this limit, surfaces degenerate to trees (or branched polymers
in the other terms). Different aspects of branched polymer physics
were discussed from the viewpoint of random surfaces in ref.~\cite{Amb}.

In some sense, this model is as interesting as any other non-critical
string theory solved so far, because all of them do not have transverse
modes on the world sheet.
Moreover, except one-dimensional string theory, they do not have
a real embedding space and can be interpreted only as the interaction of
2-$d$ quantum gravity with different matter fields.
But the $d=1$ matrix model describes trees, because the world sheet has
no area by construction. This model was solved in the planar limit in
ref.~\cite{BIPZ,KazMig} and in the double scaling in \cite{1dmatmod}.
The compactified embedding space was considered in ref.~\cite{GK},
where the exact solution for the singlet sector was obtained and
interpreted as vortex-free string theory on a ring. Non-singlet sectors
were considered in refs.~\cite{GK2,BK}.

Our mean field approach is a direct generalization
of the one of refs.~\cite{BIPZ,KazMig}. In the Hamiltonian formalism,
free fermions interacting through an effective potential appear naturally
and the Thomas-Fermi approximation is applicable.
We consider only the planar
limit, where the connection between KM and BL models holds and the mean field
approximation is exact. In the continuum limit the Hamiltonian approach
should be equivalent to the Lagrangian one. The latter has been
intensively explored in the induced QCD framework.
Saddle point equations for the KM model were derived in
ref.~\cite{Mig} and investigated in refs.~\cite{Mig,Mig2}.
The exact solution for the quadratic potential was found
in ref.~\cite{Gross}.

Mean field
gives an exact answer usually for $d$ bigger than some critical dimension
 and matrix models should not
be an exception. It means that, in principle, we can use the BL matrix model
as a convenient representation of multi-dimensional non-critical
string theory.

The outline of the present paper is the following.
In Section~2, we establish the connection
between  KM and BL models. In Section~3, we describe the mean field
approximation within the Hamiltonian formalism. To derive mean field
equations, we consider in Section~4 the external field problem
for a one-matrix integral and generalize the Brezin-Gross approach to it
\cite{BrGr}. So far only a solution for the cubic potential has been known
\cite{extfield}. Our method allows us, in principle,
to consider a general potential. In the planar limit, our approach gives
a set of equations coinciding with the Migdal's one. However, a meaning
of quantities involved is slightly different and, for finite
$N$, we still have rather simple linear equations
convenient for the $1/N$ expansion.
In Section~5, we investigate the continuum limit of the KM model.
Some problems are discussed in Section~6.
An explicit solution to the external field problem for the inverted W
potential is given in Appendix.

\section{The KM model and Bethe lattices}

Let us start from a general matrix model with an embedding space being an
arbitrary graph $\cal G$ defined by its incidence matrix $G_{xy}$
($G_{xy}=1$, if there is a link connecting
vertices $x$ and $y$, and 0, otherwise). The matrix model action reads

\be
S=-N \sum_{xy}G_{xy}tr(\Phi(x)\Phi(y))+N\sum_x tr\: V_0(\Phi(x))
\label{action}
\ee
where $\Phi(x)$ is an $N\times N$ hermitian matrix attached to an $x$'th
vertex; the potential, $V_0(\lambda)$, is an arbitrary polynomial.
The partition function is defined as the integral over all field configurations

\be
Z=\int \prod_{x\in \cal G} d\Phi(x)\; e^{-S}
\label{parfun}
\ee
In the planar limit $(N\to \infty)$, we adjust coefficients of $V_0(\lambda)$
so that to find a leading singularity of $Z$; universal continuum behavior
takes place in its vicinity.
At each vertex, $\Phi(x)$ can be decomposed into diagonal $\varphi(x)$
and angular $S(x)\in U(N)$ parts:

\be
\Phi(x)=S^+(x)\varphi(x)S(x)
\label{decomp}
\ee
It is convenient to introduce the on-link gauge variables

\be
\Omega_{xy}=S^+(x)S(y)
\label{variable}
\ee
Obviously they obey the constraint that, for every closed loop $L$,

\be
\prod_{(xy)\in L}\Omega_{xy}=\Omega_L=I
\label{constr}
\ee
where the product runs along $L$.

If the graph $\cal G$ is a tree, all gauge variables are independent
and can be integrated out by the Itzykson-Zuber (IZ)
formula \cite{ItZub}:

\be
I(\phi,\psi)=
\int  d\Omega \: e^{Ntr\phi \Omega\psi \Omega^+} = N^{-N(N-1)/2}
\prod_{n=1}^{N-1}n!\frac{\det_{ab}
e^{ N\phi_a\psi_b}}{\Delta(\phi)\Delta(\psi)}
\label{ItZub}
\ee

\ni
In eq. (\ref{ItZub}), $\phi$ and $\psi$ without loss of generality are
real and diagonal; $\Delta(\phi)=\prod_{i<j}(\phi_i-\phi_j)$
is the Van-der-Monde determinant.

As a result, one gets a model where a role of dynamical variables is
played by $N$ eigenvalues of $\Phi(x)$.
In the $N\to \infty$ limit, the mean field approximation is applicable and,
in this sense, the model is soluble.

If the graph $\cal G$ has loops, constraints (\ref{constr}) can be imposed
by averaging a number of $\delta$-functions with the corresponding
KM action:

\be
Z=Z_{\hbox{}_{KM}}
\Big\langle\; \prod_{\{L\}}\delta(\prod_{(xy)\in L}\Omega_{xy},I)
\;\Big\rangle_{\hbox{}_{KM}}
\label{Z1}
\ee
where $Z_{\hbox{}_{KM}}$ is the KM partition function for $\cal G$.
More precisely,
we treat variables (\ref{variable}) as ordinary gauge field, defined on
a given lattice $\cal G$. To recover the original matrix model, this field
has to be a pure gauge. However, eq. (\ref{Z1}) does not coincide
with the weak coupling limit of a lattice gauge partition function,
because the latter is singular and we have to impose the constraints along
non-contractible loops as well (if $\cal G$ is not simply connected as
a $d$-dimensional lattice).

Expanding $\delta$-functions in eq. (\ref{Z1})
in representations, we obtain the following integral at each link

\be
I_{ab}^R(\phi,\psi)=\int  dU \:D^R_{ab}(U)\: e^{Ntr\phi U\psi U^+}
\label{int}
\ee
where $D^R_{ab}(U)$ is a matrix element of a $U(N)$ irrep $R$;
lower indices run over a representation space $V_R$: $a,b=1,\ldots,d_R$
($d_R$ is the dimension of $V_R$); $\phi$ and $\psi$ are real and diagonal.

The integral in eq.~(\ref{ItZub})
goes actually over the right/left coset
\linebreak $U(1)^{\otimes N}\backslash
U(N)/U(1)^{\otimes N}$ rather than over the U(N) Haar measure, since
the action in eq. (\ref{ItZub}) is invariant under left and right shifts by
diagonal unitary matrices

\be
U_{ab}\to U_{ab}e^{i(\alpha_a+\beta_b)}
\label{trans}
\ee

As was shown in ref. \cite{BK}, the symmetry (\ref{trans}) gives rise
to the selection rule
for the integral (\ref{int}):

\be
I^R_{ab}(\phi,\psi)=0,{\rm \ \ unless \ \ }
\sum_{k=1}^N m_k = 0
\label{selrule}
\ee

\ni
where $m_k,\ k=1,\ldots,N$, are highest-weight components of an irrep $R$
(for $U(N)$ they are unrestricted).
Actually, as was noticed for the KM model in ref. \cite{Kogan},
in order to obtain eq.
(\ref{selrule}), it is sufficient to consider only transformations
from the center
of $SU(N)$.

Inside representation spaces, $V_R$, we also have the selection rule

\be
I^R_{ab}(\phi,\psi)=0,{\rm \ \ unless \ \ }
a,b\in V^{(0)}_R
\label{selrule2}
\ee
where $V^{(0)}_R$ is the subspace of $V_R$ spanned by all zero-weight vectors
\cite{BK}.
Its dimension is equal to

\be
d^{(0)}_R
=\int_{0}^{2\pi}\prod_{k=1}^N\frac{d\alpha_k}{2\pi}
\chi_{\scs{R}}(e^{i\alpha}),
\label{dim0}
\ee
where $\chi_{\scs{R}}(e^{i\alpha})$ is the character of an irrep $R$.
$d^{(0)}_R\neq 0$ only for irreps obeying eq. (\ref{selrule}).
When $N\to \infty$,
\be
d^{(0)}_R=0(\sqrt{d_R})
\label{asymp}
\ee

It should be stressed that the selection rule (\ref{selrule2}) cannot be
derived from the $Z_N$ symmetry of the action in (\ref{int}), because
the center acts onto representation spaces globally. The general approach
to the calculation of integrals~(\ref{int}) was recently given
in ref.~\cite{Shat}.

We can look at the $d$-dimensional KM model also from another point of
view. The absence of conditions on $\Omega$-variables is reminiscent
to matrix models with tree-like embedding spaces. We can consider a
Bethe lattice, {\em i.e.}, an infinite tree having the same coordination number
for all vertices. This lattice is a covering of a regular
lattice. We can obtain the KM model
identifying eigenvalues $\varphi(x)$ which cover the same vertices.
The $U(N)$ integration includes the sum
over all permutations and it makes eigenvalues indistinguishable.
But it is sufficient to demand the equality of corresponding momenta
of their distributions. These constraints can be interpreted as conditions on
correlators in the BL model. Actually, we restrict only
a small part
of local degrees of freedom. It suggests that, in the $N\to \infty$
limit, the KM model describes trees.
Indeed, the mean field approximation is exact in this limit in both
models and they are indistinguishable within it as far as their partition
functions are concerned.

This analogy holds also for correlators. Let us consider, for example,
the following one

\[
\Big\langle\; {\rm tr}\big( \Phi(x_1)\Phi(x_2)\big)
{\rm tr}\big(\Phi(y_1)\Phi(y_2)\big)
\;\Big\rangle_{\hbox{}_{KM}}=
\]\be
\Big\langle\;
{\rm tr}\big( \varphi(x_1)\Omega_{L_{x_1x_2}}\varphi(x_2)
\Omega_{L_{x_2x_1}}\big)
{\rm tr}\big(\varphi(y_1)\Omega_{L_{y_1y_2}}\varphi(y_2)
\Omega_{L_{y_2y_1}}\big)
\;\Big\rangle_{\hbox{}_{KM}}
\label{corr}
\ee
In the KM model we have to connect eigenvalues at different vertices by
paths $L_{z_1z_2}$ and multiply $\Omega$-matrices along them:

\be
\Omega_{L_{z_1z_2}}=\prod_{(uv)\in L_{z_1z_2}}\Omega_{uv}
\ee
Owing to the selection rule (\ref{selrule}), all appearing loops
encircle zero area, because the first non-trivial irrep
obeying eq.~(\ref{selrule}) is the adjoint. In the induced QCD
framework, this phenomenon is known as the local confinement.
If there are links through which paths from different traces go,
we can obtain a non-trivial answer also for more general loop
configurations (an example is shown in Fig.~1).
However, because of the second selection rule (\ref{selrule2}) and
eq.~(\ref{asymp}),
it has subleading order in $N$. In general, averaging $\Omega$-matrices,
we always lose half powers of $N$, and a correlator has a correct order in
$N$ only if the corresponding index loops are pure backtrackings.
Therefore, in the planar limit, the KM
model is identical to the BL one.

The ambiguity of the choice of paths can be interpreted as follows.
As far as planar graphs are concerned, the KM model is defined actually on
a covering of a regular lattice, where it is unambiguous
and all possible configurations
are tree-like.

In next orders in $N$, we lose this interpretation,
as those two models are not equivalent anymore. The geometrical reason
is clear. Already a torus has non-contractible loops which can wrap
around plaquettes without creating an embedding space area.

\section{Mean field in the Hamiltonian approach}

Let us consider the Hamiltonian formulation of the KM model.
It means that we treat one of dimensions separately as a time.
In this respect, it is a direct generalization of the standard
approach to the $d=1$ matrix model \cite{BIPZ,KazMig}.

We consider the partition function (\ref{parfun}) with the action
(\ref{action}). After the change of variables (\ref{decomp}),
(\ref{variable}) let us drop constraints (\ref{constr}) and then
integrate over $\Omega$-matrices in the $t$-direction only.
If we introduce the continuous time keeping a lattice structure in ``space
directions'', $N$ eigenvalues at each site of the $(d-1)$-dimensional
lattice become fermions
in complete analogy with refs. \cite{BIPZ,KazMig}.
We have a fermionic
ground state and, in the $N\to \infty$ limit, the Thomas-Fermi
approximation is exact. Each fermion, $\varphi_k$, moves in an effective
potential induced by its neighbours

\be
U(\varphi_k)=V_0(\varphi_k)-2(d-1)\log I[\varphi,\psi](\varphi_k)
|_{\rho(\psi)=\rho(\varphi)}
\label{effpotent}
\ee
where  $I[\varphi,\psi](\varphi_k)|_{\rho(\psi)=\rho(\varphi)}$
means the IZ integral as a function of a $k$'th eigenvalue $\varphi_k$ when
both densities coincide: $\rho(\psi)=\rho(\varphi)$.
As we are looking for the ground state in the planar limit, densities
of eigenvalues can be taken homogeneous in the space. Fermions at
different sites interact only through a collective field, since the IZ
integral depends only on traces of powers of $\varphi$ and $\psi$
separately. Hence, we have no Fermi sphere and can consider fermions at
each site independently. The one-particle Hamiltonian reads

\be
H(p,x)=\frac{1}{2}p^2+U(x)
\ee
$p$ and $x$ are a momentum and a coordinate, respectively. Now, we can
repeat standard steps \cite{BIPZ}. The Fermi level, $E_\F$,
is defined by the
equation

\be
N=\int \frac{dp\: dx}{2\pi} \;\theta(E_\F-H(p,x))
\label{N}
\ee
where $\theta(x)$ is the step function.

The ground state energy is given by the formula

\be
F=\int \frac{dp\: dx}{2\pi}\; H(p,x)\: \theta(E_\F-H(p,x))
\label{energy}
\ee

Differentiating eq.~(\ref{N}) with respect to $E_\F$, we find the
inverse frequency of a particle moving in the effective potential

\be
\frac{\partial N}{\partial E_\F}=\frac{1}{\omega(E_\F)}=\int
\frac{dp\: dx}{2\pi} \;\delta(E_\F-H(p,x))
\ee

A simple calculation gives the local density of eigenvalues

\be
\rho(x)=\frac{\int^{\hbox{}^{E_\F}}_{\hbox{}_{U(x)}}\frac{dE}{2\pi}\;
\frac{1}{\sqrt{2E-2U(x)}}}
{\int^{\hbox{}^{E_\F}}_{\hbox{}_{E_{min}}}\frac{dE}{\omega(E)}}=
\frac{1}{2\pi N}\sqrt{2E_\F-2U(x)}
\label{density}
\ee

Critical behavior occurs when the Fermi level reaches a local maximum
of the effective potential $U(x)$, which is, in general, quadratic:

\be
U(x)=U(x_o)-\frac{a}{2}(x-x_o)^2+\ldots
\label{top}
\ee
However, in our case, there is a possibility to get an answer different
from the $d=1$ model, as the coefficient $a$ can be a function of
the Fermi level $E_\F$.

Let us introduce the analytic function

\be
f(x)=\int dy\frac{\rho(y)}{x-y}
\label{f(x)}
\ee
where the integral goes over a support of $\rho(y)$. From the
one-matrix-model solution we know that, in order to get a given density
of eigenvalues, $\frac{1}{\pi}\Im f(x)$, we have to take
 $2\int \Re f(x)$ as the
one-matrix potential. The IZ interactions between lattice sites
perturb this simple solution.

To obtain a closed set of equations,
we have to express $U(x)$ through $\rho(x)$. One of possible ways to do it is
to consider the one-matrix model
in external field.

\section{External field problem revised}

Let us consider the matrix integral with the linear source term:

\be
{\cal I}[X]=\int d^{\scs{N^2}}Y \exp\, Ntr\big\{XY-V(Y)\big\}
\label{I[X]}
\ee
\ni
where $X$ and $Y$ are $N\times N$ hermitian matrices, and $V(Y)$ is an
arbitrary function defined by its Taylor expansion
$V(y)=\sum v_k y^k$.

It is obvious that ${\cal I}[X]$ depends only on
eigenvalues of $X$. The standard way to deal with the integral
(\ref{I[X]}) is to write down the Schwinger-Dyson equation

\be
\int d^{\scs{N^2}}Y {\rm tr}\Big[X-V'(Y)\Big]\exp\, Ntr\big\{XY-V(Y)\big\}=0
\label{SDeq}
\ee
\ni
and, then, rewrite it in terms of eigenvalues of $X$.

It is convenient to introduce the resolvent

\be
G_{ij}(z)=
\int d^{\scs{N^2}}Y\Big(\frac{1}{z-Y}\Big)_{ij}
\exp\, N\, {\rm tr}\big\{XY-V(Y)\big\}
\label{resolv}
\ee
\ni
Then, eq. (\ref{SDeq}) can be rewritten as follows

\be
\oint \frac{dz}{2\pi i}{\rm tr}\, \Big[\big(X-V'(z)\big)G(z)\Big]=0
\label{Eq1}
\ee
\ni
The integral goes along a small circle around $z=0$.

Eq. (\ref{Eq1}) has to be accompanied by an equation for $G(z)$ which
can be obtained from the following obvious identity

\be
{\cal I}[X]=\int d^{\scs N^2}Y\frac{1}{N}{\rm tr}\Big\{
\Big(z-\frac{1}{N}\frac{\partial}{\partial X^\dagger}\Big)
\Big(\frac{1}{z-Y}\Big)\Big\}e^{Ntr(XY-V(Y))}
\label{eq9}
\ee

Let us take $X$ almost diagonal

\be
X=x+i[A,x]+\ldots
\ee
\ni
where $A$ is an infinitesimal hermitian matrix: $\| A\| \ll 1$.
$[\ ,\ ]$ is the commutator. Then, eq. (\ref{eq9}) can be rewritten in
the form

\[
{\cal I}[X]=\frac{1}{N}\sum_{i,j=1}^{N}\bigg\{
\Big(z-\frac{1}{N}\frac{\partial}{\partial x_i}\Big)\delta_{i,j}
+\frac{1}{N}\frac{1}{x_i-x_j}\frac{\partial}{i\partial A_{ij}}
\bigg\}\]\[
\int d^{\scs N^2}Y\; \Big(\Big(\frac{1}{z-Y}\Big)_{ij}
+\Big[iA,\frac{1}{z-Y}\Big]_{ij}\Big)\,
e^{Ntr(xY-V(Y))}\Big\vert_{A=0}
\]\be =\frac{1}{N}\sum_{i=1}^{N}
\bigg\{
\Big(z-\frac{1}{N}\frac{\partial}{\partial x_i}\Big)
g_i(z)-\frac{1}{N}\sum_{j\neq i}\frac{1}{x_i-x_j}(g_i(z)-g_j(z))\bigg\}
\label{Eq2}
\ee
\ni
where

\be
g_i(z)=\int d^{\scs N^2}Y \Big(\frac{1}{z-Y}\Big)_{ii}\,
e^{Ntr(xY-V(Y))}
\label{g(z)}
\ee
are diagonal elements of the resolvent matrix $G_{ij}(z)$.
If we normalize them as

\be
{\cal W}_k(z)=\frac{g_k(z)}{{\cal I}[X]}
\ee
\ni
then we find the system of equations

\be
x_k=\oint\frac{dz}{2\pi i}V'(z){\cal W}_k(z)
\label{eqW1}
\ee
\be
1=\Big(z-\frac{1}{N}\frac{\partial}{\partial x_k}\Big){\cal W}_k(z)
-{\cal W}_k(z)\frac{1}{N}\frac{\partial}{\partial x_k}\log {\cal I}[X]
-\sum_{j\neq k}\frac{{\cal W}_k(z)-{\cal W}_j(z)}{x_k-x_j}
\label{eqW2}
\ee

Eq. (\ref{eqW2}) gives the following Laurent expansion

\be
{\cal W}_k(z)=\frac{1}{z}+
\frac{1}{z^2}\frac{1}{N}\frac{\partial}{\partial x_k}\log {\cal I}[X]
+O\Big(\frac{1}{z^3}\Big)
\ee
\ni
and, hence, we can recover ${\cal I}[X]$ from a residue of
$z{\cal W}_k(z)$.

Eqs. (\ref{eqW1}), (\ref{eqW2}) are convenient for the large $N$ limit to be
taken and also provide a suitable framework for the $1/N$ expansion.

Let us suppose that, in the $N\to\infty$ limit, there exists the density of
eigenvalues  having a finite support

\be
\rho(x)=\lim_{{\scs N}\to\infty}\frac{1}{N}\sum_{k=1}^{N}\delta(x-x_k)
\ee
\ni
Following ref. \cite{BrGr}, we can introduce a continuous
variable $x$ instead of the lower index of the function ${\cal W}_k(z)$:
$w(z,x)$. Then, in the $N\to \infty$ limit,  eqs.
(\ref{eqW1}), (\ref{eqW2})
take the form

\be
x=\oint\frac{dz}{2\pi i}V'(z)w(z,x)
\label{eqw1}
\ee
\be
1=(z-w_1(x))w(z,x)-\int dy\rho(y)\frac{w(z,x)-w(z,y)}{x-y}
\label{eqw2}
\ee
\ni
where

\be
w_1(x)=\frac{1}{N}\frac{d}{dx}\frac{\delta}{\delta \rho(x)}\log {\cal I}[X]
\ee
is the second coefficient in the Laurent expansion

\be
w(z,x)=\sum_{k=0}^{\infty}\frac{w_k(x)}{z^{k+1}}
\ee
\ni
It should be noted that eq.~(\ref{eqw2}) coincides with the one obtained
in ref.~\cite{Mig} in a slightly different context.

These equations are valid, strictly speaking, only on a support
of $\rho(x)$. Let us suppose, following ref. \cite{BrGr}, that they
can be regarded as dispersion relations and, in such a way, can be
given a meaning on the whole complex plane.

Let us introduce the function

\be
F(z,x)=\int dy\frac{\rho(y)w(z,y)}{x-y}
=\sum_{k=0}^{\infty}\frac{f_k(x)}{z^{k+1}}
\ee
\ni
where the integral runs over a support of $\rho(x)$. $F(z,x)$ is
analytic at the infinity and

\be
\Im F(z,x)=\frac{1}{\pi}\rho(x)w(z,x)
\ee

In terms of residues, eq.~(\ref{eqw2}) can be written as follows

\be
w_{k+1}(x)=(w_1(x)+f_0(x))w_k(x)-f_k(x)
\label{chain}
\ee

The functions $f_k(x)$ are not known (except $f_0(x)$, which is
determined by $\rho(x)$). However, if the potential $V(z)$ is a
polynomial, we have a finite number of equations and a bootstrap
solution can be,
in principle,
found, provided an analytic structure of all functions is guessed. No
further information is needed. In this respect, our method is a direct
generalization of the Brezin-Gross technique \cite{BrGr}.

A singularity of $w_1(x)$ at the infinity is determined by the equation
\linebreak
$V'(w_1(x))=x$. Therefore, for potentials of a degree bigger than 5,
this singularity is not algebraic and the problem is very complicated.
However, for the inverted W potential, $w_1(x)$ can be found
in a closed form by using the well-known solution to the cubic equation.
For reader's convenience we give it in Appendix.

\section{Continuum limit and scaling at $d>1$}

In the case under consideration, the density of eigenvalues of external
field coincides with the one of the integrand and we have the following
equality

\be
\int dx\;\rho(x)\, w(z,x)=\int dy\frac{\rho(y)}{z-y}
\label{cond}
\ee

In ref.~\cite{Mig} Migdal solved eq.~(\ref{eqw2}) for $w(z,x)$ after
expressing $w_1(x)$ from the Lagrangian saddle-point equation

\be
w_1(x)=\frac{1}{2d} V'_0(x)-\frac{1}{d}\Re f(x)
\label{SPE}
\ee
Then, the condition (\ref{cond}) gave the following Master Field
Equation

\be
\Re f(x) = \pvint\: \frac{dy}{\pi}\: \Im \log\Big[x-\frac{1}{2}W'(y)
+i\,\Im f(y)\Big]
\label{MFE}
\ee
where the integral goes over a support of $\Im f(y)$;
$W'(y)$ is the derivative of the effective potential

\be
U(x)=V_0(x)-2(d-1)\int dx\: w_1(x)
\ee
which, after the substitution (\ref{SPE}) for $w_1(x)$, takes the form

\be
W(x)=\frac{1}{d}V_0(x)+\frac{2(d-1)}{d}\int \Re f(x)\: dx
\label{W}
\ee

In the Hamiltonian approach we easily obtain the square root form of the
density (\ref{density}), while in the Lagrangian one the simple
saddle-point equation (\ref{SPE}) holds. Both approaches are compatible
in the continuum limit. In the simplest, d=1, case it was
demonstrated in ref.~\cite{Gross}. Let us use the same technique also for
$d>1$.

To take the continuum limit, we have to rescale the field and the
potential

\be
x=\lambda\varepsilon^{d/2-1}; \hspace{2cm}
V_0(x)=\lambda^2d\varepsilon^{d-2}+\varepsilon^d\,v(\lambda)
\ee
where $\varepsilon\to 0$ is a lattice spacing. To preserve the
normalization, we define

\be
f(x) = \varepsilon^{1-d/2} \varphi(\lambda)
\ee
In terms of new variables, eq.~(\ref{MFE}) takes the form

\be
\Re \varphi(\lambda) = \frac{1}{\varepsilon^{2-d}}\pvint \frac{d\mu}{\pi}
\Im \log\Big[\lambda-\mu-\frac{\varepsilon^2}{2d}v'(\mu)
-\frac{d-1}{d}\varepsilon^{2-d}\Re \varphi(\mu)
+i\varepsilon^{2-d}\,\Im \varphi(\mu)\Big]
\label{MFEloc}
\ee

In the $\varepsilon\to 0$ (or $d\to\infty$) limit, the term
$\varepsilon^2 v'(\mu)$ seems to be negligible with respect to
$\varepsilon^{2-d}\Re \varphi(\mu)$. However,
for the sensible continuum limit to exist, $\Re \varphi(\lambda)$ should
become $0(\varepsilon^d)$ dynamically.

Let us expand the r.h.s. of eq.~(\ref{MFEloc})
in powers of $\varepsilon$,

\[
\Re \varphi(\lambda) = \pvint \frac{d\mu}{\pi}
\frac{\Im \varphi(\mu)}{\lambda-\mu}\bigg\{
1+\frac{1}{\lambda-\mu}\bigg(\frac{\varepsilon^2}{2d}v'(\mu)
+\frac{d-1}{d}\varepsilon^{2-d}\Re \varphi(\mu)\bigg)
\]\be
-\varepsilon^{2(2-d)}\frac{[\Im \varphi(\mu)]^2}{3(\lambda-\mu)^2}
+\ldots\bigg\}
\label{expan}
\ee

as well as $\varphi(\lambda)$

\[
\Re \varphi(\lambda) = \varepsilon^{\kappa_1} r_1(\lambda) +
\varepsilon^{\kappa_2} r_2(\lambda) + \ldots
\]\be
\Im\varphi(\lambda) = \varepsilon^{\kappa_1} j_1(\lambda) + \ldots
\ee
Substituting these expansions in eq.~(\ref{expan}), we find

\[
\kappa_1=d\hspace{1cm}r_1(\lambda)=-\frac{v'(\lambda)}{2(d-1)}
\]\be
\kappa_2=d+2\hspace{1cm}j_1(\lambda)=\sqrt{\frac{4(d-1)}{d}
\int d\lambda r_2(\lambda)}
\ee
in obvious agreement with eq.~(\ref{density}).

The leading term gives the equation for $r_2(\lambda)$

\be
-\frac{v'(\lambda)}{2(d-1)} = \pvint \frac{d\mu}{\pi}
\frac{\sqrt{\frac{4(d-1)}{d}
\int d\mu r_2(\mu)}}{\lambda-\mu}
\label{udeq}
\ee
It has the form of the one-matrix-model saddle-point equation with the
upside-down potential. It is tempting to connect such a strange
phenomenon with the appearance of a tachyon in multidimensional string
theories. Its technical reason is clear: the effective potential tunes
itself to cancel a bare one up to higher order terms. Though this branch
corresponds formally to the continuum limit, it seems to describe
an effective zero-dimensional system having pathological properties.
It is unstable for most potentials, when the initial lattice model is
obviously well-defined.

If we keep only a quadratic term in the potential,
$v(\mu)=\frac{m^2}{2}\mu^2$, the solution is known from
ref.~\cite{Gross} to be

\be
\varphi(\lambda)=\frac{c\lambda}{2} -
\sqrt{\frac{c^2\lambda^2}{4}-c}
\label{gaussol}
\ee

In the $\varepsilon\to 0$ limit, there are two possible values of $c$

\be
c_1 = -\frac{\varepsilon^d m^2}{d-1}
\ee
and

\be
c_2 = \frac{4d(d-1)}{2d-1}\varepsilon^{d-2}
\label{c1}
\ee

The latter, $c_2$, does not correspond to the continuum limit.
However, this branch is always stable. If eq.~(\ref{udeq})
has no solution, the model should become gaussian in the $\varepsilon\to
0$ limit regardless of a form of the bare potential.
Of course, one could interpret this saddle-point
as a non-trivial background and construct an $\varepsilon$ expansion around it.
But it is doubtful that this would have connection with continuum
theory. This solution corresponds to the strong coupling phase of the lattice
model. In the case of a quadratic potential it was discussed by Gross
\cite{Gross}.

%%%%%%%%%%%%%%%%%%%%%%%%%%%%%%%%%%%%%%%%%%%%%%%%%%%%%%%%%%%%%%%%%%%%%%%%

However, there is some hope to obtain non-trivial behavior at least in
the Hamiltonian approach for $d\gg 1$.
A possible critical scaling could correspond here to the ``upside-down''
effective potential (\ref{top}) in which case

\be
\Im \varphi(\lambda) \sim \sqrt{-e+a\lambda^2}
\ee

In the scaling limit, $e\to 0$, we have

\be
\Re \varphi(\lambda) = 2\pvint^{\Lambda}_{\hbox{}_{\sqrt{e/a}}}
d\mu\frac{\sqrt{-e+a\mu^2}}{\lambda - \mu}
\cong -\lambda\sqrt{a}
\log \Big(\frac{\widetilde{\Lambda}^2}{e}\Big)
+ 0(1)
\ee
where we have introduced the cutoff $\widetilde{\Lambda}=2\Lambda\sqrt{a}$.

In the $d\to\infty$ limit, the presence of the continuous time should
not influence an effective potential which is proportional (as follows
from eq.~(\ref{W})) to $2\int\Re f(x)$. If we assume that the quadratic
top (\ref{top}) is really created by non-linearities at large distances,
then, from the form of the effective potential, we find the
following selfconsistency equation

\be
a \cong 2\sqrt{a}\varepsilon^{d}
\log \Big(\frac{\widetilde{\Lambda}^2}{e}\Big)
\ee
from which it follows that

\be
a \cong \bigg(2\varepsilon^{d}
\log \Big(\frac{\widetilde{\Lambda}^2}{e}\Big)\bigg)^2
\label{alog}
\ee
up to log-log terms.

Now, we are in a position to calculate the singularity of
$\frac{1}{\omega(E_\F)}$. From eq.~(\ref{top}) we have
$e\sim U(x_o)-E_\F$
and, hence,

\be
\frac{\partial}{\partial e}\Big[\frac{1}{\omega(E_\F)}\Big]\sim
\frac{\partial}{\partial e}
\int^{\Lambda}_{\hbox{}_{\sqrt{e/a}}}dx\frac{1}{\sqrt{-e+ax^2}}\sim
\frac{1}{e|\log e|}
\ee

Finally, we find that

\be
\frac{1}{\omega(E_\F)}\sim \log|\log e|
\label{loglog}
\ee

Repeating standard steps \cite{KazMig}, we can introduce a coupling
constant, $g$, by rescaling $x$ and find

\be
\frac{\partial F}{\partial E_\F}
=\frac{E_\F}{\omega (E_\F)}=E_\F\frac{\partial g}{\partial E_\F}
\ee
A critical value, $g_c$, corresponds to $E_\F$ touching the top
of the potential. Then, we get for the second derivative of the free
energy with respect to the coupling constant the following formula

\be
F''\sim \frac{1}{\log |\log \delta g|}
\ee
where $\delta g=g_c-g$. Let us remind that we have assumed that $d\gg 1$.

Unfortunately, we have no real derivation of this result, and it should
be regarded only as a reasonable proposal. If this possibility fails, it
will presumably mean that, in the continuum limit,
the KM model is effectively gaussian
(if we do not take eq.~(\ref{udeq}) seriously).

For usual branched polymers, the second derivative of the free energy
has a square root singularity, which corresponds to the string
susceptibility exponent $\gamma_{str}=1/2$ \cite{Amb}. In our case
$\gamma_{str}=0$. The reason of it could be that random surfaces have
internal degrees of freedom which produce some weights at branching
points of polymers. The double logarithm should not be surprising, as for
random surfaces $\gamma_{str}\leq 0$ \cite{Amb}, and, at $d=1$, we
have already had $F''\sim 1/\log \delta g$.

%%%%%%%%%%%%%%%%%%%%%%%%%%%%%%%%%%%%%%%%%%%%%%%%%%%%%%%%%%%%%%%%%%%%%

\section{Discussion}

The Bethe lattices are considered in statistical mechanics as simple
systems for which mean field gives exact answers. In a sense, they
correspond to the limit $d\to\infty$, {\em i.e.}, their effective
dimension is infinite. It is known that, for local $\phi^4$-type
interactions (for example), this
regime is exact actually for all dimensions bigger than 4. Presumably,
it should be the case for matrix models as well. If it is really so, the
obtained mean field solution is exact for all dimensions above a
critical one and large dimensional string theory indeed describes
branched polymers as was argued in ref.~\cite{Amb}. For the KM model
itself, the critical dimension is obviously equal to 1. If it
is independent of a tension, non-trivial string
theory does not exist.

However, the matrix models do not contain a bare string
tension as a parameter. If we substitute the heat kernel

\be
K_\beta(\Omega)=\sum_R d_R e^{-\beta C_{R}}
\chi_{\hbox{}_R}(\Omega)
\ee
for $\delta$-functions in eq.~(\ref{Z1})
($C_R$ is a second Casimir), we obtain a model
interpolating between the matrix and KM models\footnote{Another
proposal can be found in ref.~\cite{Rus}.}. Though we have no clear
interpretation for intermediate values of $\beta$, it may
bear some properties of a tension dependent matrix model. Of course, its
dynamics is extremely complicated, as we have a hybrid of a matrix model
and lattice QCD. But, if QCD string really exists, it could be used to
introduce a bare string tension in a matrix model. And, from this point
of view, the interpretation of $\beta$ as a tension may be
quite reasonable.

To conclude, we should say that the mean field problem for matrix models
is still far from being well understood. Our attempt to construct a
non-trivial continuum limit for the KM model at $d>1$ has obviously failed.
At least, eq.~(\ref{udeq}) does not seem to be good for it. On the other
hand, eq.~(\ref{loglog}), though  looks rather natural,
was not rigorously derived. It is even not clear
whether it really corresponds to a physical solution of the
model. However, in order to prove (or disapprove)
it, we would have to solve eq.~(\ref{MFEloc}) with a non-trivial
potential, which does not look to be a simple task.
Also an interpretation of the KM model in subleading orders
in $N$ remains to be done. Even for the $d=1$ model compactified on a circle,
a clear understanding of non-singlet sectors is still absent. Let us hope
that subsequent works will clarify all these problems.

\section{Acknowledgments}

I would like to thank J.Ambj\o rn, E.Brezin, V.Kazakov and I.Kostov
for the discussions
and A.A.Migdal and B.Rusakov for e-mail correspondence.
I also thank the Centre de
Physique Theorique de l'Ecole Polytechnique, where a part of this work was
done,
for hospitality.
Financial support from the EEC grant CS1-D430-C is gratefully
acknowledged.

\appendix

\section{Appendix}

We consider eq.~(\ref{I[X]}) with $V(y)$ substituted by

\be
V(y)=\frac{a}{2}y^2-\frac{1}{4}y^4
\ee
Then eq.~(\ref{eqw1}) takes the form

\be
w_1^3(x)+2f_0(x)w_1^2(x)-(a+f_1(x)-f_0^2(x))w_1(x)-f_2(x)-f_0(x)f_1(x)=x
\label{A2}
\ee
A solution to this equation can be written in the form

\be
w_1(x)=r^{1/3}(x)u(x)+\frac{v(x)}{r^{1/3}(x)}-\frac{2}{3}f_0(x)
\ee
where

\be
r(x)=\frac{x}{2}+i\sqrt{\Big(\frac{a}{3}\Big)^3
-\Big(\frac{x}{2}\Big)^2}
\ee
and $u(x)$ and $v(x)$ have the following asymptotics

\be
u(x)=1+0(1/x)\hspace{1.5cm} v(x)=\frac{a}{3}+0(1/x)
\ee
{}From the condition that $\Im w_1(x)=0$, when $\Im f_0(x)\neq 0$,
we find the only possible real solution

\be
w_1(x)=r^{1/3}(x)+\frac{a}{3r^{1/3}(x)}+\frac{1}{3}\int dy
\frac{\rho(y)}{x-y} \Big[\Big(\frac{r(x)}{r(y)}\Big)^{\frac{1}{3}} +
\Big(\frac{r(y)}{r(x)}\Big)^{\frac{1}{3}} - 2 \Big]
\ee
By substituting this ansatz in eq.~(\ref{A2}), one can check after a tedious
algebra that the former obeys the latter only on a support of $\rho(x)$.
For a cubic potential, the corresponding equation was satisfied on the
whole complex plane \cite{extfield}. The other peculiarity of
the quartic solution is that
we have no additional equation for a position of the cut.

Integrating over $x$ we find the formula for the integral (\ref{I[X]})
in the planar limit

\[
\lim_{N\to\infty} \frac{1}{N^2} \log{\cal I}[X]\; = \int dx\: \rho(x)
\bigg[\:\frac{3}{4}\Big(r^{\hbox{}^{4/3}}(x) + \Big(\frac{a}{3}\Big)^4
r^{\hbox{}^{-4/3}}(x)\Big) +
\]\[
\frac{a}{2}\Big(r^{\hbox{}^{2/3}}(x) + \Big(\frac{a}{3}\Big)^2
r^{\hbox{}^{-2/3}}(x)\Big) \bigg]
\; + \; \frac{1}{2}\int dx\: \rho(x) \int dy\: \rho(y)
\]\[
\bigg\{\log\bigg[\bigg(\frac{r(x)}{r(y)}\bigg)^{\frac{1}{3}} +
\bigg(\frac{r(y)}{r(x)}\bigg)^{\frac{1}{3}} + 1 \bigg] +
\log\bigg[\frac{3}{a}\Big(r(x)r(y)\Big)^{1/3} +
\frac{a}{3}\Big(r(x)r(y)\Big)^{-1/3} + 1 \bigg]
\]\be
- \frac{1}{2}\bigg[\bigg(\frac{r(x)}{r(y)}\bigg)^{\frac{1}{3}} +
\bigg(\frac{r(y)}{r(x)}\bigg)^{\frac{1}{3}}  +
\frac{3}{a}\Big(r(x)r(y)\Big)^{1/3} +
\frac{a}{3}\Big(r(x)r(y)\Big)^{-1/3}\bigg]\bigg\}
\ee

\end{document}